# Stray Field Magnetic Resonance Tomography using Ferromagnetic Spheres

Mladen Barbic*

Department of Physics and Astronomy, California State University, Long Beach
1250 Bellflower Boulevard, Long Beach, CA 90840

and

Axel Scherer

Department of Electrical Engineering, California Institute of Technology
1200 E. California Boulevard M/C 200-36, Pasadena, CA 91125

## ABSTRACT

The methodology for obtaining two- and three-dimensional magnetic resonance images by using azimuthally symmetric dipolar magnetic fields from ferromagnetic spheres is described. We utilize the symmetric property of a geometric sphere in the presence of a large externally applied magnetic field to demonstrate that a complete two- or three-dimensional structured rendering of a sample can be obtained without the motion of the sample relative to the sphere. Sequential positioning of the integrated sample-sphere system in an external magnetic field at various angular orientations provides all the required imaging slices for successful computerized tomographic image reconstruction. The elimination of the requirement to scan the sample relative to the ferromagnetic tip in this imaging protocol is a potentially valuable simplification compared to previous scanning probe magnetic resonance imaging proposals.

* Electronic mail: mbarbic@csulb.edu

There has been a steady advance in the field of Magnetic Resonance Imaging (MRI) towards higher resolution, with the ultimate goal of atomic imaging capability. The largest measurement challenges stem from weak signals typical of high-resolution magnetic resonance [1], and the limitation of available gradient field strengths from current carrying conductors. Following the original reports [2-4] of applying magnetic field gradients to samples in order to demonstrate magnetic resonance imaging of spatial spin distribution, improvements in conventional inductive detection [5,6] have resulted in spatial imaging resolution of approximately $1\mu m$ [7-10]. The attraction and intense research interest towards 3D MRI with higher resolution is driven by the well-known advantages of MRI as a three-dimensional, non-invasive, multi-contrast, and chemically specific imaging tool [11,12].

The introduction of ferromagnetic nanostructures for increased sensitivity and resolution in magnetic resonance imaging has opened additional avenues toward achieving the atomic resolution goal. Scaling considerations show that a miniaturized permanent magnet will produce higher fields than an electromagnet, and can be further scaled to a smaller size without any loss in field strength [13]. Miniaturization of permanent magnets also provides an increase in the magnetic field gradients while requiring no electrical power supply and no current leads. Finally, permanent magnets generate no heat and thus require no heat dissipation. This ability of nanometer scale ferromagnets to provide ultra-high magnetic field gradients that can in turn spatially resolve resonant spins on the atomic scale has led Sidles to propose the Magnetic Resonance Force Microscope (MRFM) [14]. In this instrument, a microscopic magnetic particle on a mechanical cantilever acts as a source of atomic scale imaging gradient fields as well as a force generator on the spins whose magnetic resonance the mechanical cantilever detects [15]. Magnetic resonance image can be obtained by mechanically scanning the tip in three dimensions over the sample.

In this article, we focus on the magnetic resonance imaging protocol that uses the interaction between a sample and the stray fields from a geometrically symmetric ferromagnetic sphere. We demonstrate that a two- or three-dimensional imaging of a sample can be obtained without the motion of the sample relative to the sphere. We believe that the use of such a sphere model is reasonable, as microscopic ferromagnetic spheres of high quality and shape uniformity have been successfully fabricated, manipulated, and integrated on sensors [16-19]. Our protocol is similar to the Stray Field Magnetic Resonance Imaging (STRAFI) technique [20] where constant magnetic field gradients, on the order of 60T/m, from superconducting magnets are used. Here, however, the nanometer scale ferromagnetic spheres can provide ultra-high magnetic field gradients ($\sim$5x10$^6$T/m for a 100nm diameter Cobalt sphere), that can in principle be utilized for three-dimensional magnetic resonance imaging with resolution reaching Angstrom levels.

We describe a two-dimensional imaging protocol first, before expanding this principle to the full three-dimensional method. Our model configuration is shown in Figure 1a, where a sample with size of $\sim$1/10 the size of the ferromagnetic sphere is positioned as shown. The sample can represent a biological cell $\sim$10$\mu$m in size next to a sphere 100$\mu$m in diameter, or at the extreme, a small molecule or protein of $\sim$10nm in size next to a 100nm diameter ferromagnetic sphere. A large DC magnetic field $B_0$ ($\sim$10 Tesla) is applied parallel to the z-direction, polarizing the spins of the sample as well as saturating the magnetization of the ferromagnetic sphere. A small radio frequency field $B_1$ is applied perpendicular to the large polarizing DC magnetic field $B_0$. In the absence of the ferromagnetic sphere, the nuclear spins in the sample would experience the same externally applied field $B_0$ and therefore meet the magnetic resonance condition at the same magnetic resonance frequency $\omega_R$. However, close to the ferromagnetic sphere, a large magnetic field gradient is present at the sample, and only certain spins of the sample satisfy the correct magnetic resonance condition at any given magnetic field and frequency:

$$\omega(\vec{r}) = \gamma \left| B(\vec{r}) \right| \qquad (1)$$

The magnetic field from the ferromagnetic sphere at point *r* in the sample has the following azimuthally symmetric dipolar form:

$$\vec{B}(\vec{r}) = \frac{3\vec{n}(\vec{m}\cdot\vec{n}) - \vec{m}}{\left|\vec{r}\right|^3} \qquad (2)$$

where *n* is the unit vector that points from the center of the ferromagnetic sphere to the sample location, and *m* is the magnetic moment vector of the sphere. Since the external DC polarizing magnetic field $B_0$ is considered to be much larger than the field from the ferromagnetic sphere, only the z-component of the magnetic field from the ferromagnetic sphere, $B_Z$, needs to be considered [21-23] for imaging. For a ferromagnetic sphere, this z-component of the magnetic field has the azimuthally symmetric form:

$$B_Z(\vec{r}) = \frac{M_0}{\left|\vec{r}\right|^3}(3\cos^2\theta - 1) \qquad (3)$$

where $\theta$ is the angle between the z-axis and the distance vector *r*, and *M₀* is the magnitude of the saturation magnetic moment of the ferromagnetic sphere. Figure 1(a) also shows the contours of constant values for the z-component of the magnetic field from the sphere, $B_Z$, along the x-z plane.

In contrast to the previous approaches [21-23], we propose to fix the sample directly on the sphere, as shown in Figure 1(a), at an angular location where:

$$\frac{\partial B_Z(\vec{r})}{\partial r} = \frac{M_0}{|\vec{r}|^4}(3\cos^2\theta - 1) = 0 \qquad (4)$$

At this angular orientation of $\theta=54.7°$, $B_Z\approx 0$, and the contours of constant z-component of the magnetic field $B_Z$ from the ferromagnetic sphere are perpendicular to the sphere surface, so that the sample is intersected by approximately perpendicular imaging slices. In Figure 1(b), the contours of constant z-component of the magnetic field from the ferromagnetic sphere are shown along the plane parallel to the two-dimensional sample surface. This view shows that the magnetic resonance spectrum of the two-dimensional sample (i.e., the configuration shown in Figure 1) will be a one-dimensional projection of the sample spin density. This leads to the possibility of obtaining a computerized tomographic image [24-27] if multiple imaging slices from the dipolar field of the ferromagnetic sphere can be obtained at different angles, as we describe below.

The imaging slices at multiple angles required for the computerized tomographic image reconstruction process can be obtained from a configuration of Figure 1 without the motion of the sample relative to the sphere. We come to this conclusion by considering what happens when the integrated sample/sphere system is jointly rotated by an angle $\varphi$ around the $\theta=54.7°$ axis, as shown in Figure 1. Although both the sample and the sphere are mechanically rotated by the same angle $\varphi$, the presence of a large polarizing magnetic field $B_0$ of ~10 Tesla along the z-axis ensures that the saturated magnetic moment of the ferromagnetic sphere remains oriented along the z-axis. As a result, the imaging contours of constant z-component of the magnetic field, $B_Z$, remain fixed in space. Therefore, rotating the fixed sample/sphere system at a uniform sequence of angles φ provides all of the required imaging slices for previously developed two-dimensional computerized tomography reconstruction algorithms [24-27].

We note that, depending on the instrumental constraints or preferences, the actual rotation of the integrated ferromagnetic-sphere/sample system shown in Figure 1 could also be experimentally executed by multiple sequential rotations around the x and y axes, as shown in Figure 2. As rotations do not commute, such sequential rotations around x- and y-axes would have to be carefully selected. For example, the rotation of the sample around $\theta_Y$ and then around $\theta_X$, shown in Figure 2a, would result in the correct translation and rotation of the sample for proper tomographic slicing, while a single rotation around the z-axis, shown in Figure 2b, would result in the correct translation but incorrect rotation of the sample for proper slicing by the contours of constant $B_Z$. Additionally, we restrict our sample size to a fraction of the ferromagnetic sphere dimension in order to maintain the slicing of the sample by approximately parallel contours of constant $B_Z$. We note that image reconstruction from non-parallel slices has been demonstrated in computerized tomography [26] and is mathematically justified [28,29].

In order to extend our methodology to the three-dimensional imaging case, we find it advantageous to represent the integrated sphere/sample system rotations (described in Figures 1 and 2) in a precessing ferromagnetic sphere moment reference frame, as shown in Figure 3. In this perspective, although much harder to implement experimentally for a $B_0$=10 Tesla magnetic field, the same effect of image slicing as described in Figures 1 and 2 can be employed. In this reference frame, the sample is fixed and located on top of the sphere, as shown in Figure 3, while the ferromagnetic moment of the sphere is tilted away from the z-axis by $\theta=54.7°$ and precessed around the z-axis at a sequence of angles $\varphi$ required for the tomographic image reconstruction process.

We now analyze the case of a three-dimensional sample mounted on a ferromagnetic sphere, as shown in Figure 4. At the angular position of $\theta=54.7°$, as in the two-dimensional imaging case, the sample is intersected by the planes of constant z component of the magnetic field from the ferromagnetic sphere that are approximately perpendicular to the sphere surface. Consider now the rotation of the integrated sample/sphere system so that the angle $\varphi=0$ is held fixed while the angle $\theta$ is sequentially reduced in value from $\theta=54.7°$ to $\theta=0$. This results in the sequential slicing of the three-dimensional sample by the imaging planes that range from being approximately perpendicular to the sphere surface to being approximately parallel to the sphere surface, as Figure 4(a) shows. Therefore, by rotating the sample/sphere system through several angular values that range from $\theta=54.7°$ to $\theta=0$, all the required imaging slices are obtained for two-dimensional image reconstruction along the x-z plane where angle $\varphi=0$. This protocol again relies on the principle that, although both the sample and the sphere are mechanically rotated by the angle $\theta$, the large polarizing magnetic field along the z-axis ensures that the saturated magnetic moment of the ferromagnetic sphere remains oriented along the z-axis and the imaging contours remain fixed in space.

A three-dimensional imaging protocol follows directly from these principles as all of the slices needed for three-dimensional image reconstruction can be obtained by varying both angles $\varphi$ and $\theta$, as described in the precessing ferromagnetic sphere moment reference frame of Figure 4(b). By sequentially varying the angles (θ, φ) of the ferromagnetic moment direction through all possible angular combinations from θ=54.7˚ to θ=0˚ and φ=0˚ to φ=360˚, as shown in Figure 4(b), the sample will be intersected by imaging slices at all possible angular orientations. This is sufficient for a complete three-dimensional image reconstruction, although several points of interest need to be addressed regarding the image-reconstruction process.

It is apparent from Figure 4(a) that the planes of constant z-component of the dipolar magnetic field $B_Z$ from the ferromagnetic sphere are curved, non-parallel, and not equally spaced. This is not prohibitive for the image reconstruction procedure, as basic back-projection algorithms [24-27] can be used for obtaining a three-dimensional image of the sample. More specifically, for an angular orientation (θ, φ), a weighted value is assigned to each contour of constant z-component of the magnetic field $B_Z$ from the magnetic resonance spectrum obtained at that angular orientation. The three-dimensional image reconstruction of the sample is then completed by repeating the weighted value assignment procedure for all angular orientations (θ, φ). Although this procedure is sufficient for basic three-dimensional image reconstruction, this simple back-projection algorithm is known to produce star-like image artifacts, and is therefore not optimal. The less artifact-prone but more complicated filtered back-projection algorithms or, alternatively, the matrix-based iterative-reconstruction algorithms could be employed [26].

A second point of interest is the image resolution. It is apparent from the inspection of the contours of constant z-component of the magnetic field in Figure 4(a) that the image resolution depends on the distance from the ferromagnetic sphere surface.

Only two magnetic field gradient forms are of interest since there is no variation of the azimuthally symmetric contours of the constant z-component of the magnetic field with the change of angle φ. The variation of the imaging contours along the radial direction is described by Equation 4, and the gradient of the imaging contours along the angular θ direction is:

$$\frac{1}{r}\frac{\partial B_Z(\vec{r})}{\partial \theta} = -\frac{M_0}{|\vec{r}|^4}(6\cos\theta\sin\theta) = -\frac{3M_0}{|\vec{r}|^4}\sin 2\theta \qquad (5)$$

Both gradients have an inverse radial dependence to the fourth power, which means that parts of the sample closer to the sphere will experience higher magnetic field gradients and therefore can in principle be imaged with a higher resolution. This can also be deduced from Figure 4(a). Strong dependence of the gradient fields on $r$ in Equations 4 and 5 also explains why the use of the nanoscale ferromagnetic spheres is advantageous in potentially obtaining atomic resolution images from projections.

It is important to point out that in our imaging method it is not required to know a priori where the sample is located on the ferromagnetic sphere. If the ferromagnetic moment direction is sequentially varied through the angles (θ, φ) from θ=0° to θ=180° and φ=0° to φ=360°, the sample will be intersected by the imaging slices at all possible angular orientations, and a three-dimensional image reconstruction through back-projection algorithms will reveal an image and the location of the sample on the ferromagnetic sphere.

In addition to understanding the imaging methodology and resolution, it is important to discuss the choice of experimental methods for sample/sphere positioning as well as magnetic resonance detection. Our protocol involves angular motion of the sample/sphere system around two rotational axes. Such sample positioning technology is well developed and is routinely used in the STRAFI technique [20]. The sensitivity requirements depend on the desired resolution. In addition to the conventional inductive detection, we also suggest that optical detection methods [30-32], micro-coils [33,34], superconducting quantum interference devices (SQUID) [35,36], Hall sensors [37,38], and superconducting resonators [39] remain viable candidates to be implemented in this imaging method. In addition, cantilever detection could be employed through direct transfer of angular momentum [40-43] and energy [44-46] to the spin population in the magnetic resonance process. As compared to the scanning probe type cantilever detection [15], in our protocol the need for scanning the sample with respect to the ferromagnetic probe is eliminated, along with the potential problems of long term positioning drift between the sample and the ferromagnetic gradient source. It is also important to note that, with the elimination of the relative motion of the sphere with respect to the sample, the thermo-mechanical vibrations of the cantilever do not translate into relative thermal motion and therefore fluctuations of the magnetic fields and field gradients from the sphere at the sample location. The intrinsic thermal motion of the magnetic moment remains, however, and has to be carefully considered in the ferromagnetic sphere material selection [47].

We have described a technique for magnetic resonance tomography using the dipolar magnetic fields from ferromagnetic spheres distinctly different from previous magnetic resonance scanning probe microscopy approaches that seek to achieve atomic imaging resolution. In previous experimental schemes, the images are obtained by raster scanning a ferromagnetic probe over the sample in three dimensions, and de-convolving intensities from the obtained magnetic resonance spectra at each point [48,49]. In contrast, in the dipolar field magnetic resonance tomography scheme described in this article, the ferromagnetic sphere and the sample are fixed with respect to one another. We rely on the geometric symmetry of the sphere and on the principle that the ferromagnetic moment remains saturated and oriented along a large polarizing magnetic field despite the mechanical motion of the sphere. Angular positioning of the integrated sample/sphere system then provides all the required imaging slices for computerized tomographic image reconstruction. The elimination of the requirement of scanning the sample relative to a ferromagnetic tip in this new imaging protocol could represent a valuable experimental simplification and bring us closer to the goal of atomic resolution in three-dimensional nuclear magnetic resonance imaging.

This material is based upon work supported by the National Science Foundation under the NSF-CAREER Award Grant No. 0349319 and by the National Institute of Health Grant NIH-RO1 HG002644. The authors thank Dr. Joyce Wong for helpful discussions and comments on the manuscript.

**References**

1. Abragam, *Principles of Nuclear Magnetism*, Oxford University Press, New York (1983).

2. P. C. Lauterbur, *Nature (London)* **242**, 190 (1973).

3. P. Mansfield and P. K. Grannell, *J. Phys. C* **6**, L422 (1973).

4. P. Mansfield and P. K. Grannell, *Phys. Rev. B* **12**, 3618 (1975).

5. D. I. Hoult and R. E. Richards, *J. Magn. Reson.* **24**, 71 (1976).

6. D. I. Hoult and P. C. Lauterbur, *J. Magn. Reson.* **34**, 425 (1979).

7. J. Aguayo, S. Blackband, J. Schoeniger, M. Mattingly, and M. Hintermann, *Nature (London)* **322**, 190 (1986).

8. S.-C. Lee, K. Kim, J. Kim, S. Lee, J. H. Yi, S. W. Kim, K.-S. Ha and C. Cheong, *J. Magn. Reson.* **150**, 207 (2001).

9. L. Ciobanu, D. A. Seeber, and C. H. Pennington, *J. Magn. Reson.* **158**, 178 (2002).

10. A. Blank, C. R. Dunnam, P.P. Borbat, and J. H. Freed, *J. Magn. Reson.* **165**, 116 (2003).

11. P. T. Callaghan, *Principles of Nuclear Magnetic Resonance Microscopy*, Oxford University Press, New York (1991).

12. B. Blumich, *NMR Imaging of Materials*, Oxford University Press, New York (2000).

13. K. Halbach, *J. Appl. Phys.* **57**, 3605 (1985).

14. J. A. Sidles, *Appl. Phys. Lett.* **58**, 2854 (1991).

15. J. A. Sidles, J. L. Garbini, K. J. Bruland, D. Rugar, O. Zuger, S. Hoen, and C. S. Yannoni, *Rev. Mod. Phys.* **67**, 249 (1995).

16. M. Barbic, J. J. Mock, A. P. Gray, and S. Schultz, *Appl. Phys. Lett.* **79**, 1897 (2001).

17. D. R. Baselt, G. U. Lee, K. M. Hansen, L. A. Chrisey, R. J. Colton, *Proc. IEEE* **85**, 672 (1997).

18. M. A. Lantz, S. P. Jarvis, and H. Tokumoto, *Appl. Phys. Lett.* **78**, 383 (2001).

19. T. Ono and M. Esashi, *Rev. Sci. Instrum.* **74**, 5141 (2003).

20. P. J. McDonald and B. Newling, *Rep. Prog. Phys.* **61**, 1441 (1998).

21. M. Barbic, *J. Appl. Phys.* **91**, 9987 (2002).

22. M. Barbic and A. Scherer, *J. Appl. Phys.* **92**, 7345 (2002).

23. M. Barbic and A. Scherer, *J. Appl. Phys.* **95**, 3598 (2004).

24. G. T. Herman, *Image Reconstruction from Projections* Academic Press, New York (1980).

25. F. Natterer, *The Mathematics of Computerized Tomography*, John Wiley & Sons, New York (1986).

26. A. C. Kak and M. Slaney, *Principles of Computerized Tomographic Imaging*, SIAM, Philadelphia (2001).

27. S. R. Deans, *The Radon Transform and Some of Its Applications*, Krieger Publishing Company, Malabar (1993).

28. A. M. Cormack, *Proc. Amer. Math. Soc.* **83**, 325 (1981).

29. A. M. Cormack, *Proc. Amer. Math. Soc.* **86**, 293 (1982).

30. J. Wrachtrup, C. Vonborczyskowski, J. Bernard, M. Orrit, and R. Brown, *Nature (London)* **363**, 244 (1993).

31. J. Kohler, J. A. J. M. Disselhorst, M. C. J. M. Donckers, E. J. J. Groenen, J. Schmidt, and W. E. Moerner, *Nature (London)* **363**, 242 (1993).

32. I. M. Savukov and M. V. Romalis, *Phys. Rev. Lett.* **94**, 123001 (2005).

33. D. L. Olson, T. L. Peck, A. G. Webb, R. L. Magin, and J. V. Sweedler, *Science* **270**, 1967 (1995).

34. A. G. Webb, *Prog. Nucl. Magn. Reson. Spectrosc.* **31**, 1 (1997).

35. L. R. Narasimhan, C. K. N. Patel, and M. B. Ketchen, *IEEE Trans. Appl. Supercond.* **9**, 3503 (1999).

36. Ya. S. Greenberg, *Rev. Mod. Phys.* **70**, 175 (1998).

37. G. Boero, P. A. Besse, and R. Popovic, *Appl. Phys. Lett.* **79**, 1498 (2001).

38. J. Jin and X-Q Li, *Appl. Phys. Lett.* **86**, 143504 (2005).

39. R. D. Black, T. A. Early, P. B. Roemer, O. M. Mueller, A. Mogro-Campero, L. G. Turner, and G. A. Johnson, *Science* **259**, 793 (1993).

40. E. Arimondo, *Il Nuovo Cimento* **52**, 8583 (1967).

41. G. Alzetta, E. Arimondo, C. Ascoli, and A. Gozzini, *Il Nuovo Cimento* **52**, 8596 (1967).

42. C. Ascoli, P. Baschieri, C. Frediani, L. Lenci, M. Martinelli, G. Alzetta, R. M. Celli, and L. Pardi, *Appl. Phys. Lett.* **69**, 3920 (1996).

43. M. Lohndorf, J. Moreland, and P. Kabos, *Appl. Phys. Lett.* **76**, 1176 (2000).

44. J. Schmidt and I. Solomon, *J. Appl. Phys.* **37**, 3719 (1966).

45. J. Moreland, M. Lohndorf, P. Kabos, R. D. McMichael, *Rev. Sci. Instrum.* **71**, 3099 (2000).

46. A. Jander, J. Moreland, and P. Kabos, *Appl. Phys. Lett.* **78**, 2348 (2001).

47. J. D. Hannay, R. W. Chantrell, and D. Rugar, *J. Appl. Phys.* **87**, 6827 (2000).

48. O. Zuger and D. Rugar, *Appl. Phys. Lett.* **63**, 2496 (1993).

49. O. Zuger, S. T. Hoen, C. S. Yannoni, and D. Rugar, *J. Appl. Phys.* **79**, 1881 (1996).

**Figure Captions**

**Fig. 1.** (a) Model configuration for two-dimensional magnetic resonance tomography using ferromagnetic spheres. Imaging contours of constant z-component of the magnetic field are perpendicular to the sphere surface and intersect the sample positioned at θ=54.7˚. (b) The imaging contours shown along the plane parallel to the two-dimensional sample surface. The magnetic resonance spectrum of the sample is a one-dimensional projection of the sample spin density. Sequential rotations by angle φ provide the required projections for the tomographic image reconstruction process.

**Fig. 2.** (a) Alternative procedure for proper image slicing by sequential rotations around the y- and x-axes. (b) A single rotation around the z-axis would result in an incorrect rotation of the sample for proper slicing by the magnetic field imaging contours.

**Fig. 3.** Precessing ferromagnetic sphere moment reference frame. The sample is fixed and located on top of the sphere while the magnetic moment of the sphere is tilted away from the z-axis by θ=54.7˚ and precesses around the z-axis at a sequence of angles φ.

**Fig. 4.** (a) Rotation of the integrated sample/sphere system from θ=54.7˚ to θ=0 results in the sequential slicing of the three-dimensional sample by the imaging planes that range from being approximately perpendicular to the sphere surface to being approximately parallel to the sphere surface. (b) The precessing magnetic moment reference frame for the three-dimensional tomography. By sequentially varying of the angles (θ, φ) of the sphere magnetic moment direction through all angular combinations from θ=54.7˚ to θ=0˚ and φ=0˚ to φ=360˚, the sample is intersected by the imaging slices at all possible angular orientations.

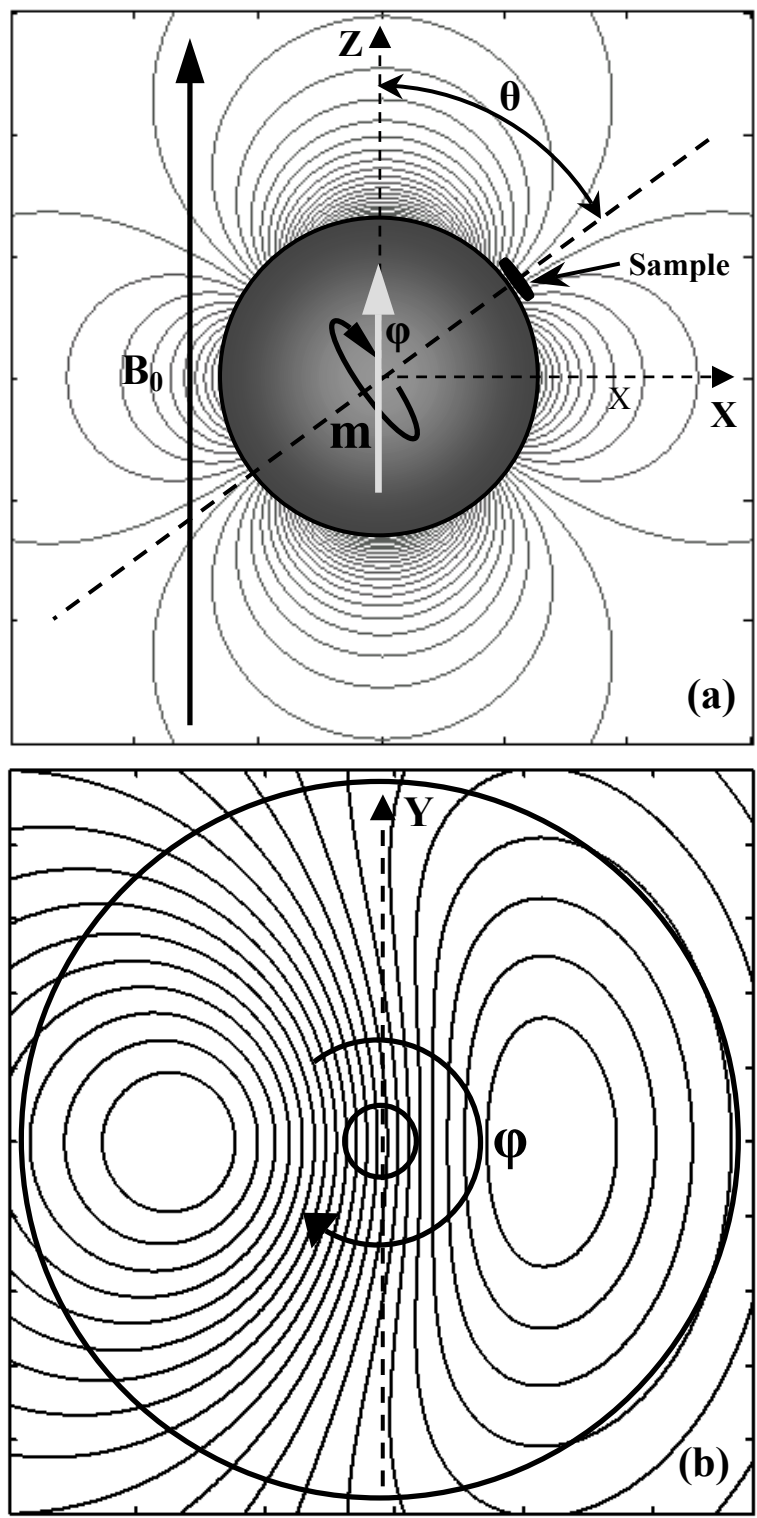

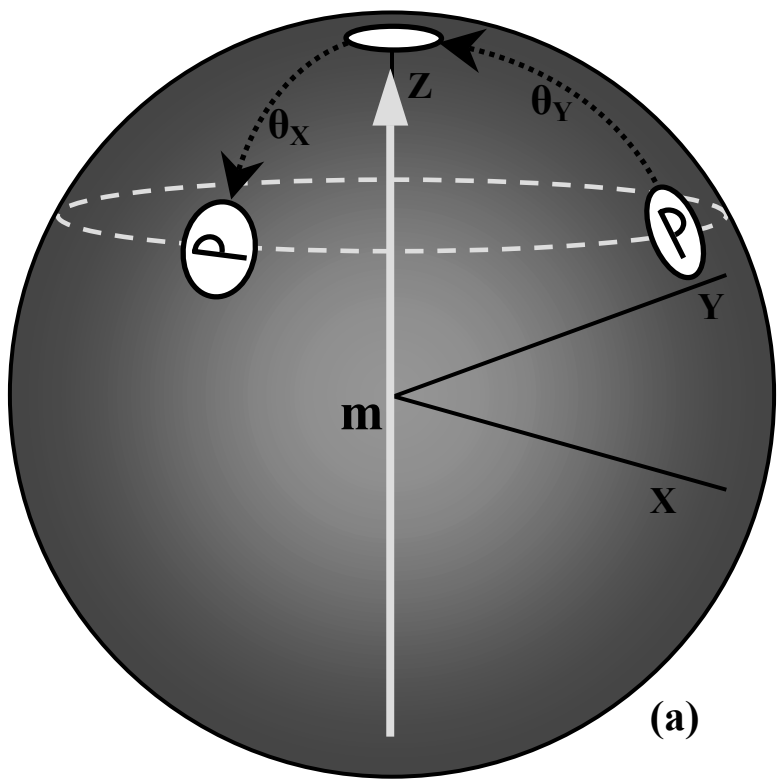

**(a)**

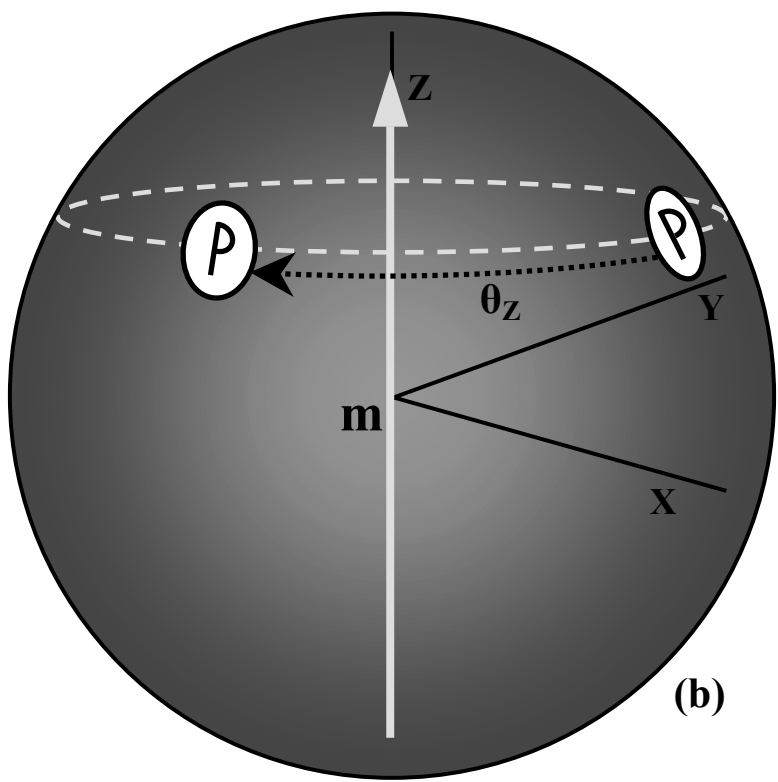

**(b)**

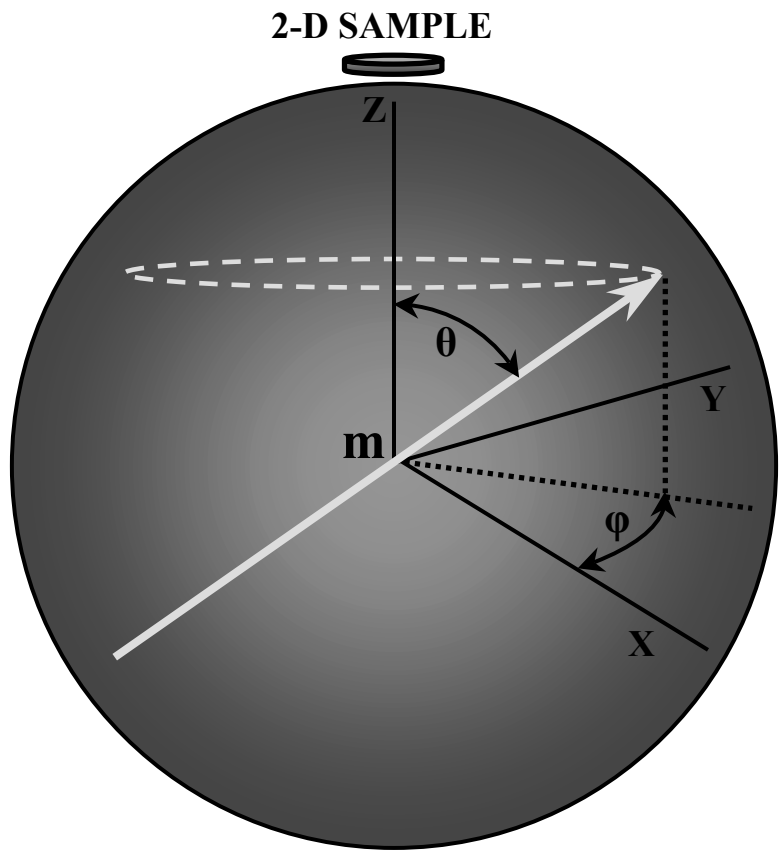

Figure 4

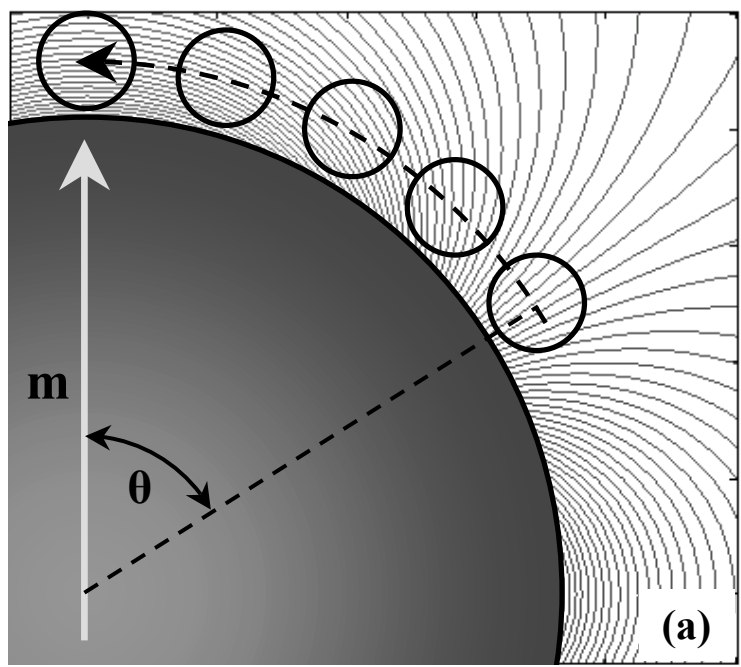

**(a)**

**3-D SAMPLE**

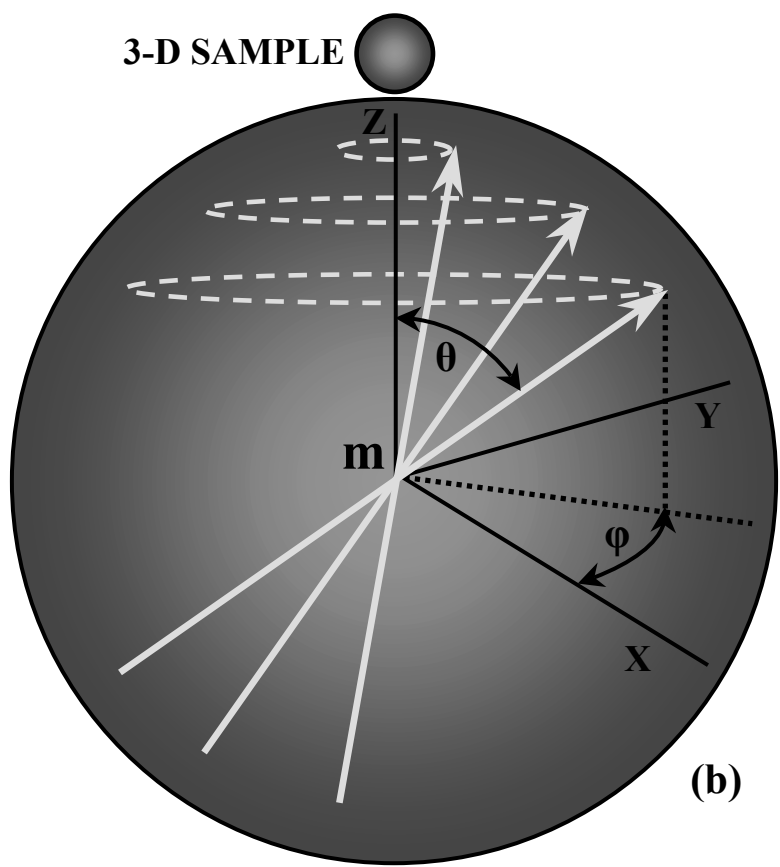

**(b)**